\documentclass{revtex4}
\usepackage{graphicx}
\usepackage{amsmath}

\begin{document}

\title{Electric/magnetic flux tube on the background of
magnetic/electric field}
\author{Vladimir Dzhunushaliev}
\email{dzhun@hotmail.kg}
\affiliation{Dept. Phys. and Microel. Engineer., Kyrgyz-Russian
Slavic University, Bishkek, Kievskaya Str. 44, 720000, Kyrgyz
Republic}

\date{\today}

\begin{abstract}
It is argued that the phenomenon of a flux tube in quantum chromodynamics 
is closely connected with a spontaneously symmetry breakdown of gauge theory. 
It is shown that in the presence of a mass term in the SU(2) gauge theory 
the Nielsen-Olesen equations describe the flux tube surrounded by an external 
field. 
\end{abstract}

\maketitle

\section{Introduction}

One of the more fascinating aspects of quantum chromodynamics (QCD) 
is a quark confinement. The phenomenon of the confinement of quarks is 
connected with a hypothesized flux tube filled with the color electric field 
and stretched between quark and antiquark. This situation is in contrast with 
the electron and positron pair where the force lines in the whole space 
are spread. The confinement is a non-perturbative effect, it 
means that it can not be derived using traditional theoretical 
methods of the quantum field theory, i.e. Feynman diagram 
technique. The main problem for the derivation 
of confinement from the first principles is the absence of quantization 
methods for such strongly nonlinear field theories as SU(3) gauge field.
The lattice simulations confirm the point of view that
in QCD there is such phenomenon as the flux tube filled with a
chromoelectric field. Another confirmation of this point of view 
is a dual QCD \cite{baker} where the electric and magnetic fields switch places 
and a dual Meissner effect with respect to the chromoelectric force lines 
takes place. In this theory there is a famous Nielsen-Olesen flux tube 
solution \cite{no} filled with longitudinal magnetic field which can imitate 
the flux tube stretched between quark and antiquark. 
\par
Evidently one of the most essential problems in the confinement phenomenon 
is the derivation of a field distribution for the flux tube
from first principles. At first, similar problem was set up by Heisenberg 
\cite{heis}: he has calculated some characteristics of a quantized nonlinear 
spinor field. His basic idea is that the fundamental equation for 
the quantized nonlinear spinor field is the
corresponding classical equation where the classical spinor field is replaced
by an operator of the spinor field. This equation is applied for the derivation
of an infinite equations set for Green's functions which give us all information 
relevant for quantized spinor field.
\par
The idea presented here is the same \cite{vdsin1}: we assume that the 
fundamental equations in QCD is Yang-Mills equations where the gauge potential 
$A^a_\mu$ is replaced by an operator $\hat{A}^a_\mu$. 
Later, according to Heisenberg, we derive equations for field correlators 
that is similar to ideas proposed in Ref. \cite{giacomo}. In the first 
approximation we will assume that 
$\left\langle Q | f(A^a_\mu) | Q \right\rangle =
f( \left\langle Q | A^a_\mu | Q \right\rangle )$
here $| Q \rangle $ is a quantum state. But this
is not all : we assume that the non-linearity of the SU(2) gauge theory leads
to symmetry breaking, i.e. to the appearance of some additional terms
in the Yang-Mills equations. We presuppose that these terms is like to
$(-m^2(a) A^a_\mu)$ (here the brackets
$\langle \ldots \rangle$ are omitted). This assumption is similar to a 
Coleman-Veinberg mechanism \cite{coleman} in $\lambda \varphi^4$ theory. 
The essence of the Coleman-Veinberg mechanism is that the nonlinear terms like 
$\varphi^4$ produce some corrections by such a way that a spontaneously 
spontaneously breakdown 
occurs. In the present letter we explicitly incorporate mass terms of some 
components of the gauge potential into the SU(2) Yang-Mills equations to 
describe dynamics of the flux tube gauge fields in the presence of an 
external field. In Ref. \cite{deguchi} similar idea is presented to derive 
an effective Abelian gauge theory of a modified SU(2) gluodynamics. 
Although the mass-generation mechanism is not well understood analytically 
at present we accept the mass generation as true in the beginning 
of our discussion without questioning its mechanism. We expect that the 
mass terms of gauge potential are dynamically induced by a non-perturbative 
effect of gluodynamics. 
\par 
In the next section we will obtain equations set \eqref{sec2-60}-\eqref{sec2-90} 
which describes a cylindrically symmetric distribution of the SU(2) gauge field 
coupled with the Higgs scalar field. These equations contain the 
\textit{abelian} Nielsen-Olesen 
flux tube : \eqref{sec2-80}-\eqref{sec2-90} equations with $f=v=0$. One can hope that 
these equations can involve another 
interesting flux tube solutions with \textit{non-abelian} gauge fields. In this note 
we do the first step in this direction: we show that these equations actually 
have solutions which can be interpreted as electric/magnetic flux tube on the 
background of magnetic/electric field. We hope that there is the next step 
on which we will obtain the more complicated flux tube solutions with 
\textit{longitudinal non-abelian chromoelectric} field which is absent in the 
Nielsen-Olesen flux tube. 
\par 
Therefore the main goal of this letter is to show that spontaneously symmetry 
breakdown of the SU(2) gauge theory coupled with the Higgs scalar field leads to 
a variant of the SU(2) flux tube filled with electric/magnetic field surrounded 
by an external magnetic/electric field. 

\section{Ansatz and equations}

The equations for the SU(2) gauge potential $A^a_\mu$ are
\begin{eqnarray}
  D_\nu F^{a\mu\nu} &=& e \epsilon^{abc} \phi^b D^\mu \phi^c
  - m^2(a) A^{a\mu} ,
\label{sec2-10}\\
  D_\mu  D^\mu \phi^a &=& -\lambda \phi^a
  \left( \phi^b \phi^b - \phi^2_\infty \right)
\label{sec2-20}
\end{eqnarray}
here $e$ and $\phi_\infty$ are some constants;
$D_\mu (\ldots)^a = \partial_\mu (\ldots)^a +
e \epsilon^{abc} A^b_\mu (\ldots)^c$ is the covariant derivative; 
$\phi^a$ is a scalar field which is by analogy with the Nielsen-Olesen 
flux tube. Let us remind that the term $m^2(a) A^a_\mu$ is the consequence 
of the quantization (which leads to a gauge symmetry breadown) 
and equations \eqref{sec2-10}, \eqref{sec2-20} are some approximation for 
the calculation of the averaged
gauge potential $\langle A^a_\mu \rangle$. We consider the cylindrically
symmetric case and consequently ansatz for the gauge potential we choose
in the form \cite{Obukhov:1996ry}
\begin{eqnarray}
  A^1_t(\rho) &=& \frac{f(\rho)}{e} ,
\label{sec2-30}\\
  A^2_z(\rho) &=& \frac{v(\rho)}{e} ,
\label{sec2-40}\\
  A^3_\varphi (\rho) &=& \frac{\rho w(\rho)}{e} ,
\label{sec2-50}\\
  \phi^1(\rho) &=& \frac{\phi(\rho)}{e} ,
\label{sec2-60}
\end{eqnarray}
here $\rho, z$ and $\varphi$ are the cylindrical coordinates. The corresponding
equations are
\begin{eqnarray}
  f'' + \frac{f'}{\rho} &=& f\left( v^2 + w^2 \right) - m^2 f ,
\label{sec2-60}\\
  v'' + \frac{v'}{\rho} &=& v\left( -f^2 + w^2 + \phi^2 \right) - m^2 v ,
\label{sec2-70}\\
  w'' + \frac{w'}{\rho} - \frac{w}{\rho^2} &=&
  w\left( -f^2 + w^2 + \phi^2 \right) ,
\label{sec2-80}\\
  \phi'' + \frac{\phi'}{\rho} &=& \phi\left( v^2 + w^2 \right) +
  \lambda \phi \left( \phi^2 - \phi^2_\infty \right) ,
\label{sec2-90}
\end{eqnarray}
here we have replaced $e\phi \rightarrow \phi$,
$\lambda/e^2 \rightarrow \lambda$,
$e \phi_\infty \rightarrow \phi_\infty$, $m(1) = m(2) = m$, $m(3) = 0$. 
The condition $m_3 = 0$ means that the U(1) gauge symmetry remains unbroken. 
Noticing that the system
\eqref{sec2-60}-\eqref{sec2-90} is close to that of the spherically symmetric
case in the Prasad-Sommerfeld limit \cite{prasad}, we look for the general
solution for $f,v,\phi$ in the form \cite{Obukhov:1996ry}
\begin{equation}
  v(\rho) = K(\rho), \; \phi(\rho) = K(\rho) \cosh \gamma , \;
  f(\rho) = K(\rho) \sinh \gamma
\label{sec2-100}
\end{equation}
here $\gamma$ is an arbitrary constant. Additionally we will consider the
limit
\begin{equation}
  \lambda \rightarrow 0, \; \text{but} \;
  \lambda \phi^2_\infty = m^2 . 
\label{sec2-110}
\end{equation}
After which we have the following equations
\begin{eqnarray}
  K'' + \frac{K'}{\rho} &=& Kw^2 + K\left( K^2 - m^2 \right) ,
\label{sec2-120}\\
  w'' + \frac{w'}{\rho} - \frac{w}{\rho^2} &=& 2w K^2 .
\label{sec2-130}
\end{eqnarray}
Let us introduce the dimensionless coordinate $x = \rho m \sqrt{2}$ 
and replace $K/m \rightarrow K$, $w/(m\sqrt{2}) \rightarrow w$ 
\begin{eqnarray}
  K'' + \frac{K'}{x} &=& Kw^2 + \frac{1}{2} K\left( K^2 - m^2 \right) ,
\label{sec2-140}\\
  w'' + \frac{w'}{x} - \frac{w}{x^2} &=& w K^2 .
\label{sec2-150}
\end{eqnarray}
This is the famous Nielsen-Olesen equations for the flux tube and with special
choice of $\lambda = 1/2$. These equations are good investigated, see for example
Ref's. \cite{n-o}. It is well known that \eqref{sec2-140} and \eqref{sec2-150} 
equations are equivalent to the first order
equations
\begin{eqnarray}
  \frac{F'}{x} +\frac{s_\pm}{2} \left( K^2 - 1 \right) &=& 0 ,
\label{sec2-160}\\
  K' + s_\pm \frac{F K}{x}&=& 0
\label{sec2-170}
\end{eqnarray}
here $F = xw$ and $s_\pm = \pm 1$. These equations are some special case 
of BPS strings which describe flux tube solutions in dual SU(N) theories 
\cite{kneipp}. 

\section{Magnetic and electric flux tubes on the background of electric
and magnetic fields}

\begin{figure}[h]
  \begin{minipage}[t]{.45\linewidth}
  \begin{center}
    \fbox{
    \includegraphics[height=5cm,width=5cm]{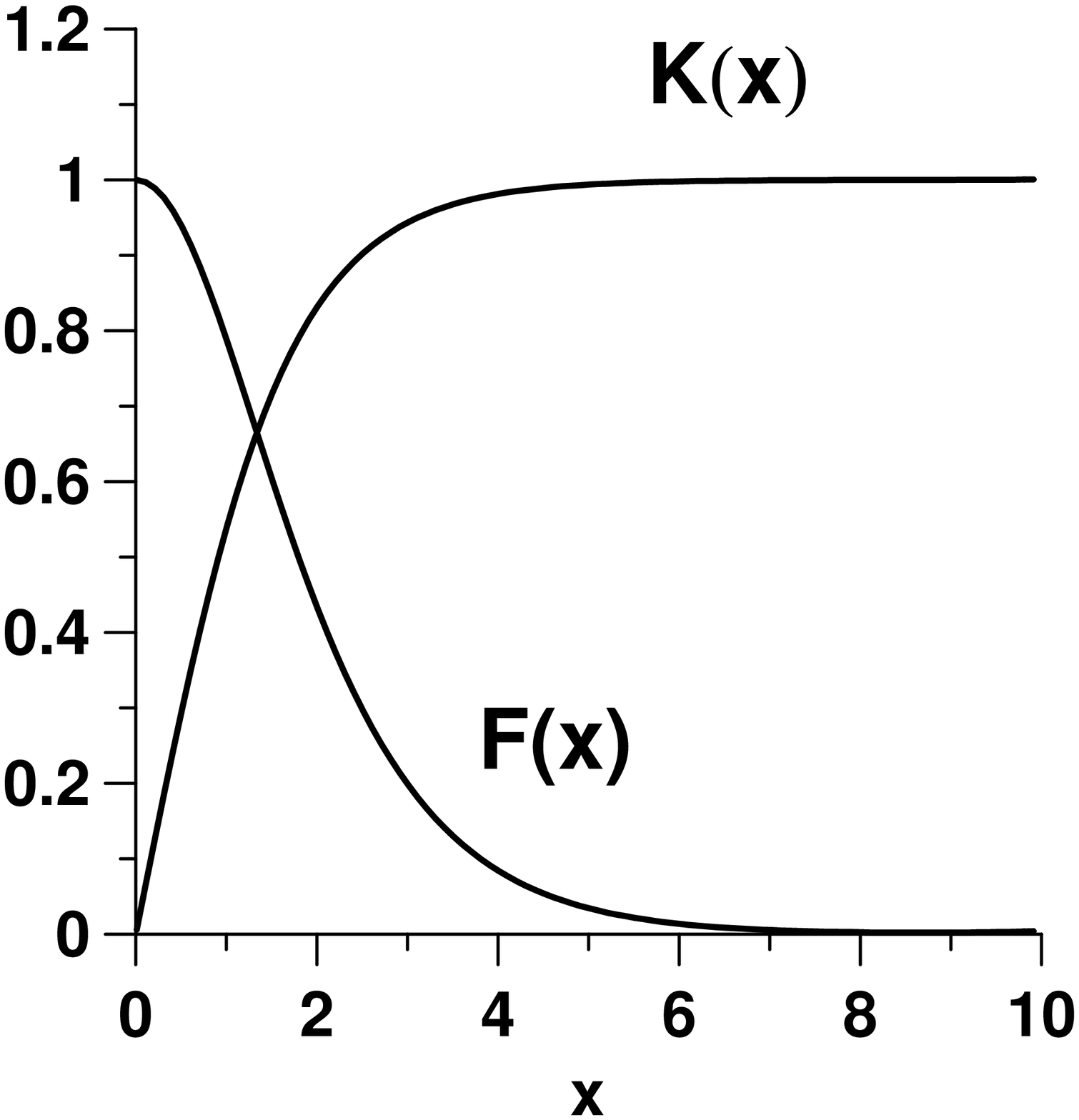}}
    \caption{The first case. The functions $F(x)$ and $K(x)$.}
    \label{fig1}
  \end{center}
  \end{minipage}\hfill
  \begin{minipage}[t]{.45\linewidth}
  \begin{center}
    \fbox{
    \includegraphics[height=5cm,width=5cm]{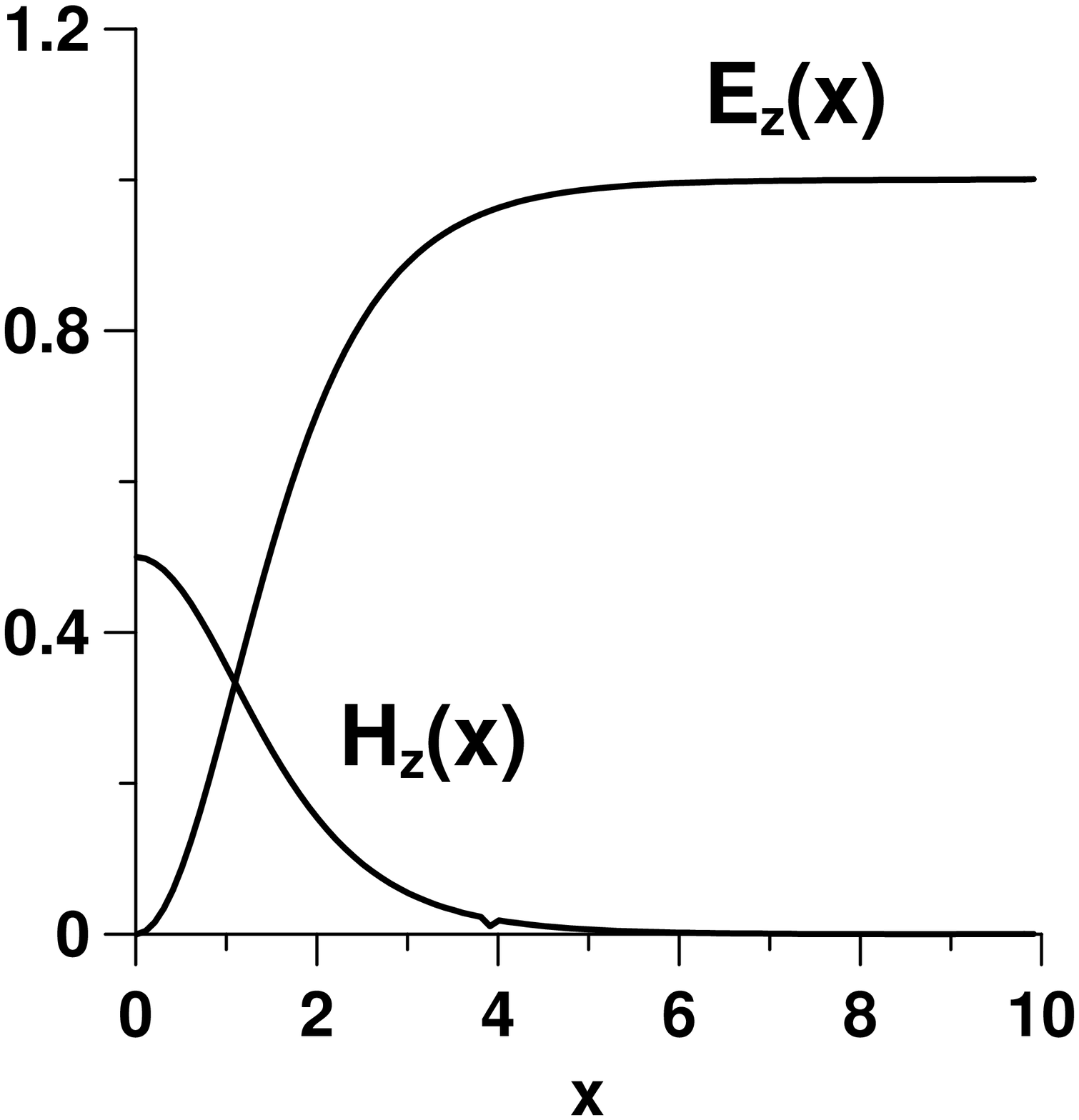}}
    \caption{The first case. The electric $E_z$ and magnetic $H_z$ 
    fields.}
    \label{fig2}
  \end{center}
  \end{minipage}
\end{figure}

It is well known that equations \eqref{sec2-160}-\eqref{sec2-170} describe
the Nielsen-Olesen flux tube. This solution at the origin is
\begin{eqnarray}
  F &=& 1 - \frac{x^2}{4} + \cdots ,
\label{sec3-10}\\
  K &=& b x + \cdots ,
\label{sec3-20}\\
  s_- &=& -1
\label{sec3-30}
\end{eqnarray}
here $b$ is an arbitrary constant. The asymptotical behavior is
\begin{eqnarray}
  F(x)\biggl |_{x \rightarrow \infty} &\rightarrow& 0 ,
\label{sec3-40}\\
  K(x)\biggl |_{x \rightarrow \infty} &\rightarrow& 1 .
\label{sec3-50}
\end{eqnarray}
This solution describes the magnetic field localized in a tube. Both functions
are presented on Fig. \ref{fig1}. The flux of magnetic
field is quantized which means that the solution exists only for a special
choice of $b \approx 0.6031985$.
\par
But our interpretation of this solution for the SU(2) gauge theory with symmetry
breaking is different. In our case we have the following color fields
\begin{eqnarray}
  E^3_z = F^3_{tz} = e \epsilon^{3bc}A^b_t A^c_z = \frac{fv}{e} &=&
  \frac{K^2 \sinh \gamma}{e} ,
\label{sec3-60}\\
  H^3_z = -\rho F^{3\rho \varphi} &=& - \frac{F'}{\rho} .
\label{sec3-70}\\
  H^2_\varphi = F^{2\rho z} = \partial_{\rho} A^2_z = \frac{v'}{e} &=&
  \frac{K'}{e} ,
\label{sec3-80}\\
  E^2_\varphi = F^{2t\varphi} = -A^1_t A^3_\varphi = -\rho f w &=& 
  -\rho w K \sinh \gamma ,
\label{sec3-90}\\
  E^1_\rho  = F^1_{\rho z} = \partial_{\rho} A^1_t = \frac{f'}{e} &=&
  \frac{K' \sinh \gamma}{e} ,
\label{sec3-93}\\
  H^1_\rho = \rho F^{1z\varphi} = vw &=& K w .
\label{sec3-99}
\end{eqnarray}
For us interesting is only $E_z$ and $H_z$ fields which are presented on
Fig.\ref{fig2}.
\par
\begin{figure}[h]
  \begin{minipage}[t]{.45\linewidth}
  \begin{center}
    \fbox{
    \includegraphics[height=5cm,width=5cm]{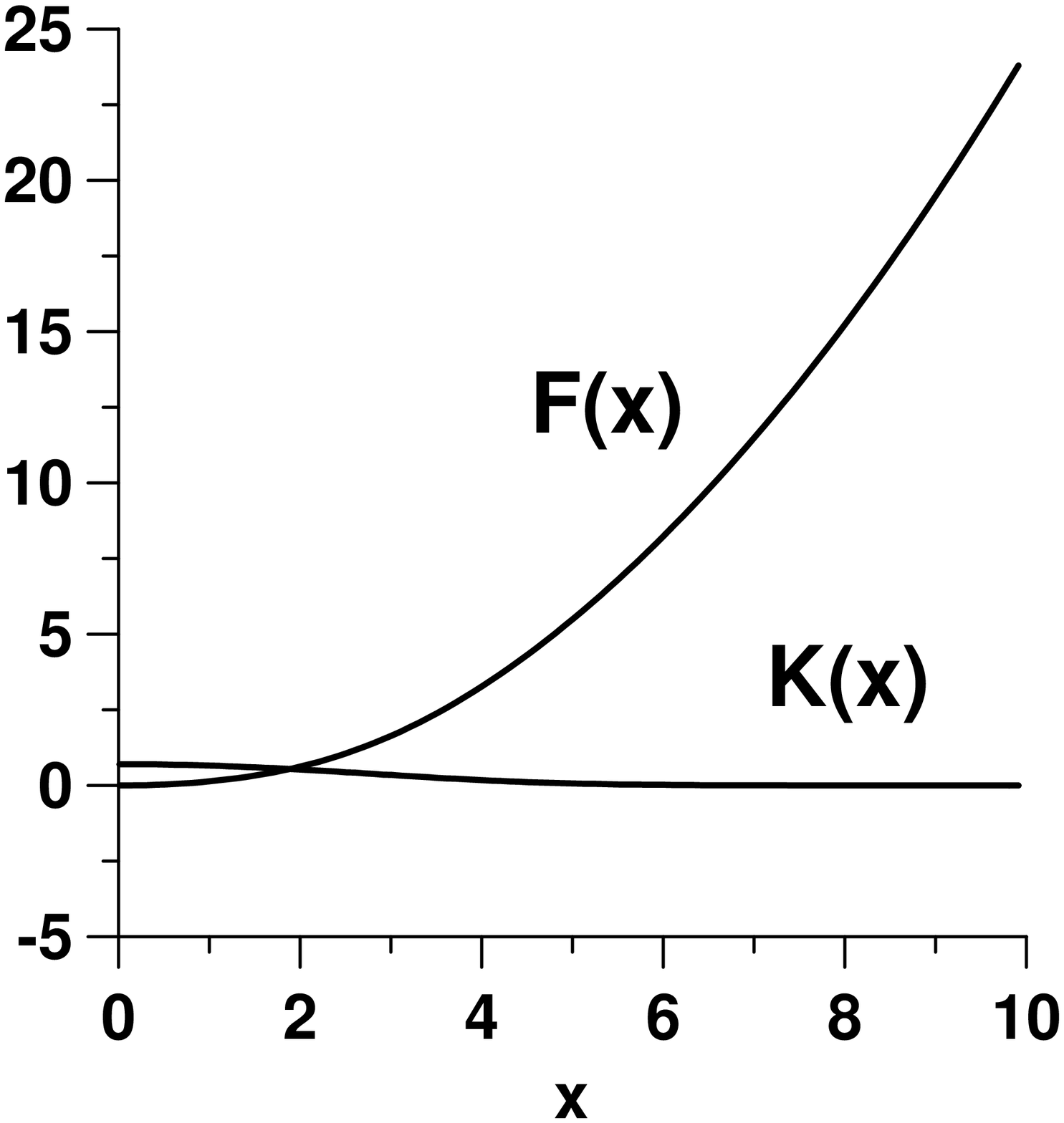}}
    \caption{The second case. The functions $F(x)$ and $K(x)$.}
    \label{fig3}
  \end{center}
  \end{minipage}\hfill
  \begin{minipage}[t]{.45\linewidth}
  \begin{center}
    \fbox{
    \includegraphics[height=5cm,width=5cm]{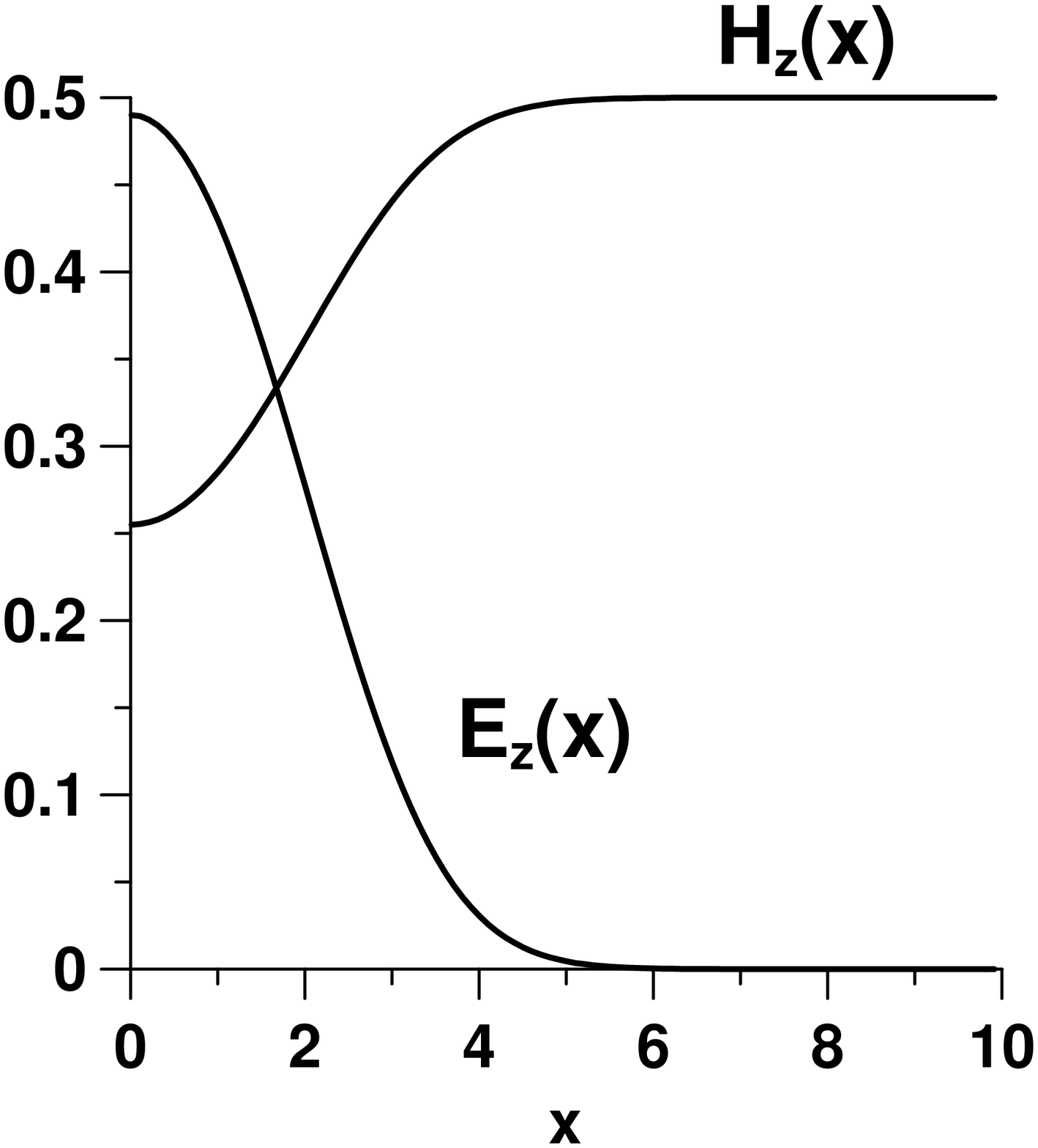}}
    \caption{The second case. The electric $E_z$ and magnetic $H_z$ 
    fields.}
    \label{fig4}
  \end{center}
  \end{minipage}
\end{figure}

The second type of solution has the following form at the origin
\begin{eqnarray}
  F &=& \frac{1}{4} \left( 1 - K_0^2 \right)x^2 + \ldots ,
\label{sec3-100}\\
  K &=& K_0 - \frac{1}{8} K_0 \left( 1 - K_0^2 \right)x^2 + \ldots ,
\label{sec3-110} \\
  s_+ &=& +1
\label{sec3-120}
\end{eqnarray}
here $K_0^2 < 1$ is an arbitrary constant. The solution has the following
asymptotical behavior
\begin{eqnarray}
  F(x)\biggl |_{x \rightarrow \infty} &\rightarrow& \frac{x^2}{4} ,
\label{sec3-130}\\
  K(x)\biggl |_{x \rightarrow \infty} &\rightarrow& K_\infty e^{-\frac{x^2}{8}} .
\label{sec3-140}
\end{eqnarray}
here $K_\infty$ is some constant. In contrast with the Nielsen-Olesen flux tube
this solution describes a flux of electric field on the background of the 
magnetic field. The solution exists for
arbitrary $K_0^2 < 1$ and this means that the flux of electric field is
non-quantized. The typical solution is presented on Fig.\ref{fig3}.
On Fig.\ref{fig4} the electric and magnetic fields are plotted. 
\par 
We see that in both cases one field (electric/magnetic) pushes out another 
one (magnetic/electric). 

\section{Discussion and conclusions}

Our interpretation of presented solutions is the following : we have obtained 
the flux tube filled with the longitudinal electric/magnetic field on the background 
of the longitudinal external magnetic/electric field. The word ''external`` 
demands an explanation. The non-Abelian gauge theories are nonlinear and 
consequently an external field must be a part of the solution of Yang-Mills 
equations which describes simultaneously the flux tube and external fields. 
Thus presented solutions is a \textit{nonlinear superposition} of the flux tube and 
external fields.
\par 
We have shown that spontaneously symmetry breakdown probably is an essential 
ingredient of a quantum gauge theory leading to a flux tube field distribution. 
The question about the mechanism of this phenomenon is very complicated. 
The origin of symmetry breakdown can be connected with the non-linearity 
of non-Abelian theories, ghost condensation \cite{dudal} or maybe even with 
properties of an algebra of field operators \cite{dzhun}.  
\par 
We have mentioned above that every electric/magnetic field in the flux tube 
pushes out the external magnetic/electric field. It can be connected with 
the fact that we group together $A^a_t, A^a_z$ components of the gauge 
potential and the scalar field $\phi^a$. From the 't Hooft-Polyakov monopole 
and Nielsen-Olesen flux tube we know that the scalar field pushes out the gauge 
fields. Thus in our case it may be that the separation of the flux tube and 
external fields is connected with the bundle \eqref{sec2-100} of some gauge 
potential components and scalar field. Probably that the breakdown of the bundle 
\eqref{sec2-100} will lead to a flux tube without any external gauge field 
and it should be the next step of the investigation in this direction. 
\par 
In closing I want to underline that this consideration is not pure classical 
one as we add the mass term which is connected with the quantization 
of such non-linear theory as the SU(2) gauge theory. 

\section{Acknowledgments}
I am very grateful to the ISTC grant KR-677 for the financial support.

\end{document}